\documentclass[
 reprint, 
 floatfix,
 superscriptaddress,
 amsmath, amssymb,
 aps,
 pre,
]{revtex4-2}

\usepackage{graphicx}% Include figure files
\usepackage{siunitx} % units
\usepackage{newtxtext,newtxmath} % new
\usepackage{makecell}

\usepackage[%
    linkcolor  = blue, 
    citecolor  = blue, 
    urlcolor   = blue, 
    colorlinks = true
]{hyperref}

\begin{document}

%\preprint{AIP/123-QED}

\title{Experimental investigation of the flow rate--pressure drop relation of a viscoelastic Boger fluid in a deformable channel}

\author{SungGyu Chun}
  \affiliation{Mechanical Science and Engineering, University of Illinois at Urbana--Champaign, Urbana, Illinois \mbox{61801, USA}}

\author{Ivan C. Christov}%
 \affiliation{School of Mechanical Engineering, Purdue University, West Lafayette, Indiana \mbox{47907, USA}}%

\author{Jie Feng}
  \email{jiefeng@illinois.edu}
  \affiliation{Mechanical Science and Engineering, University of Illinois at Urbana--Champaign, Urbana, Illinois \mbox{61801, USA}}

\date{\today}% It is always \today, today,
             %  but any date may be explicitly specified

\begin{abstract}
We study the steady flow-induced deformation between an incompressible non-Newtonian fluid and a three-dimensional (3D) deformable channel. Specifically, we provide a comprehensive experimental--theoretical framework for such flows of constant-viscosity viscoelastic (i.e., Boger) fluids, which allows us to quantify the influence of the fluid's viscoelasticity on the flow rate--pressure drop ($q-\Delta p$) relation. For a flow-rate-controlled regime and weakly viscoelastic flow, we find excellent agreement between the theoretical prediction and experimental results, specifically providing an experimental demonstration of how both wall compliance and fluid viscoelasticity decrease the pressure drop. To definitively demonstrate the coupled effect of wall compliance and fluid viscoelasticity (change of cross-sectional area and curving of streamlines), we perform experiments in an equivalent rigid channel, showing that both the Boger fluid and a viscosity-matched Newtonian fluid have the same pressure drop. Thus, the experimentally verified theory for the $q-\Delta p$ relation of weakly viscoelastic flows of Boger fluids in compliant channels can now be used for the design of microfluidics and soft hydraulic systems operating with complex fluids. 
\end{abstract}

\maketitle

\section{INTRODUCTION}

The field of soft hydraulics---characterized by the intricate interplay between flow-induced forces on the inherently compliant boundary of microscopic conduits---has gained considerable attention within the scientific community in recent years. This growing attention stems from the critical role of soft hydraulics across a wide range of applications, such as the design of organ-on-a-chip platforms \cite{bhatia2014microfluidic}, wearable electronics \cite{kashaninejad2023microfluidic,fande2025recent}, and soft robotics \cite{polygerinos2017soft,matia2023harnessing}, to name a few. In these applications, understanding the relationship between the flow rate $q$ and pressure drop $\Delta p$ in various compliant geometries is crucial for device performance and functionality. Although the flow rate--pressure drop ($q-\Delta p$) relations for the laminar flow of Newtonian fluids in common geometries are well established in textbooks, the same is not true for non-Newtonian, specifically \emph{viscoelastic}, fluids flowing through non-uniform rigid \cite{boyko2022pressure,housiadas2023lubrication,boyko2024flow,mahapatra2025viscoelastic} and deformable \cite{ramos2021fluid,boyko2023non,boyko2025interplay} conduits, despite their relevance in many practical applications such as microfluidics \cite{anna2013non,delgiudice2018fluid,battat2022nonlinear} and tribology \cite{tichy1996non,veltkamp2023lubrication,sari2024effect}. Importantly, both the rheology of the complex fluids and the flow-induced non-uniform conduit deformation lead to a \emph{nonlinear} $q-\Delta p$ relation \cite{gervais2006flow,christov2018flow,karan2020flow,christov2021soft,PhysRevFluids.9.043302,boyko2025interplay}, which is important to characterize for microfluidic applications \cite{delg2016is,suteria2018nicrofluidic,battat2022nonlinear}.
Furthermore, microfluidic techniques have been increasingly proposed and employed to measure complex fluids' rheology and specifically their viscoelasticity  \cite{pipe2009microfluidic,gupta2016microfluidic,delg2016is}. Yet, an experimental measurement of how the interplay between viscoelasticity and wall compliance sets the $q-\Delta p$ relation in steady low-Reynolds-number flow within a deformable configuration remains elusive. 

Previous investigations of the $q-\Delta p$ relations for the pressure-driven flow of viscoelastic fluids in rigid channels have predominantly relied on numerical simulations \cite{zografos2020viscoelastic, varchanis2022reduced} and experimental measurements \cite{ober2013microfluidic, james2021pressure}, providing key insights into the complex interplay between fluid elasticity and flow resistance. For deformable geometries, two-dimensional (2D) numerical simulations \cite{chakraborty2010viscoelastic,chakraborty2015viscoelastic} using the Oldroyd-B \cite{oldroyd1950formulation,hinch2021oldroyd}, finite-extensibility nonlinear elastic (FENE) type, and Owens constitutive equations have shown that the viscoelasticity of the fluid strongly influences the forces acting on a short deformable wall segment in a channel. The viscoelastic nature of a complex fluid is expected to influence the $q-\Delta p$ relations \cite{boyko2022pressure} of such flows through non-uniform geometries. Specifically, flow-induced deformation causes the streamlines of the flow to curve, induces elastic stresses, and changes the flow resistance. Viscoelastic corrections to the $q-\Delta p$ relation would thus be expected. Importantly, however, at low Deborah number $De\ll1$ (i.e., for weakly viscoelastic flow), most viscoelastic constitutive equations' predictions reduce to those of the Oldroyd-B model \cite{housiadas2023lubrication,boyko2024perspective}.  Recently, employing lubrication theory with the structural deformation of a thick plate and performing a perturbation expansion in $De\ll1$, \citet{boyko2023non} developed a theory for the $q-\Delta p$ relation under the Oldroyd-B model, taking into account the flow-induced deformation, and quantifying how both fluid viscoelasticity and channel compliance reduce the pressure drop.

On the other hand, to the best of our knowledge, there is no quantitative comparison between these theoretical predictions and experimental data for constant-viscosity viscoelastic (i.e., Boger \cite{boger1977highly,james2009boger}) fluids. We find that the lack of prior experiments is partly due to challenges in the preparation of stable working fluids, as well as the challenges of accurately measuring the pressure drop change in weakly viscoelastic flows. 
In this study, we overcome these challenges and provide an experimental demonstration of the coupled effects of fluid viscoelasticity and wall compliance, guided by quantitative comparisons between theory and experiments. We design an experimental platform to measure the pressure drop, for an imposed steady flow rate, in the canonical microfluidic configuration of a three-dimensional (3D) rectangular channel with a deformable top wall. 

\begin{figure}[ht]
    \includegraphics[width=\columnwidth]{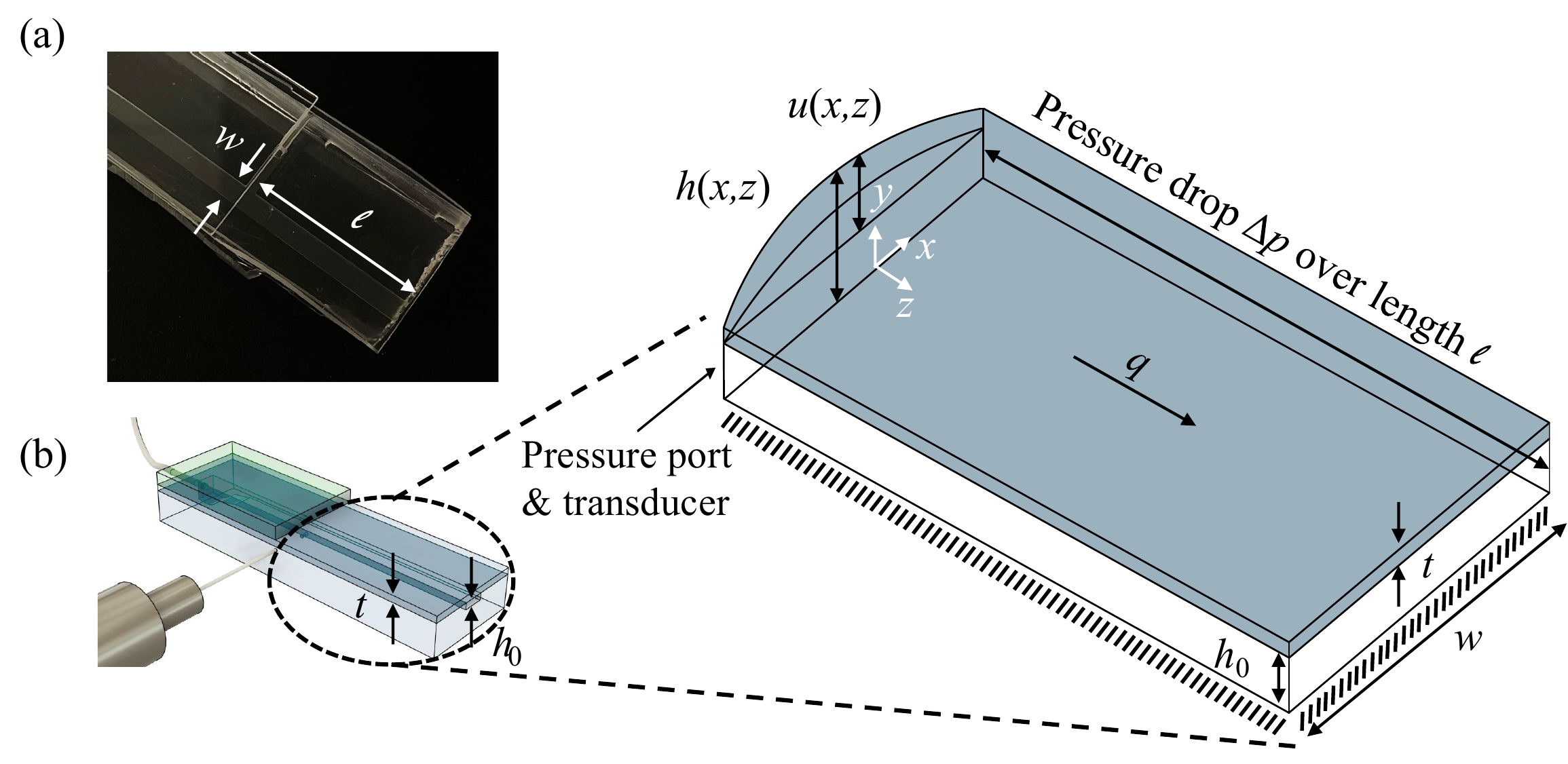}
    \caption{\label{fig1}Illustration of the configurations used for experiments and modeling. (a) Image of the experimental system and (b) schematic of the 3D rectangular channel geometry with a top deformable wall, showing the mathematical notation. The geometry consists of a compliant section of length $\ell$, with a pressure port placed along the stream-wise direction to measure the $\Delta p$ across $\ell$ between the pressure port and the outlet, which is open to the atmosphere. The initially rectangular cross-section has a width, $w$, and undeformed height, $h_0$, with $h_0\ll w \ll \ell$ (see Table~\ref{T1}). The thickness of the top deformable wall is $t$, and $u$ is the vertical displacement of the fluid--solid interface. The steady imposed flow rate is $q$.}
\end{figure}

\section{EXPERIMENTAL SETUP}

\subsection{Design and fabrication of the deformable microchannel}

We follow our previous work on shear-thinning \cite{PhysRevFluids.9.043302} in designing and fabricating the microfluidic device used in the experiments. We designed the device mold using a Computer Aided Design software (Fusion 360, Autodesk) and fabricated it using a 3D printer (ELEGOO Mars Resin 3D Printer) based on the required (slender, $h_0 \ll \ell$, and shallow, $h_0 \ll w$) channel dimensions (Fig.~\ref{fig1}). We prepared the polydimethylsiloxane elastomer (PDMS, Sylgard 184, Dow Corning) with a mixing ratio of 10:1 (w/w) between the silicone elastomer base and the curing agent. Then, we poured the PDMS mixture onto the 3D printed mold and degassed the mixture under vacuum for two hours to eliminate small air bubbles. After curing the degassed mixture in an oven at $90\si{\celsius}$ for 12 hours, we carefully removed the PDMS channel from the mold and punched holes to provide a flow inlet and a pressure sensor port (diameter $d = 1.07 \pm 0.05~\si{\milli\meter}$). Next, we treated both the PDMS channel and thin PDMS film (GASKET-500PK, SiMPore) with a 4.5 MHz hand-held corona treater (BD-20AC, Electro-Technic Products) for 30 seconds and brought them together in conformal contact for bonding. To restrict the wall deformation to the region of interest, we further attached a rigid glass slide on top of the thin PDMS film using oxygen plasma bonding, leaving only the deformable region (length $\ell$, see Fig.~\ref{fig1}) exposed. We then verified the dimensions of the fabricated channel through microscope visualization and summarized the dimensions in Table~\ref{T1}. In addition, the Young's modulus $E$ and the Poisson's ratio $\nu$ of the thin PDMS were measured using a dynamic mechanical analyzer (DMA 850, TA Instruments) at $25\si{\celsius}$ and are summarized in Table~\ref{T1}.

\begin{table*}
    \caption{\label{T1}Physical dimensions of the slender ($h_0 \ll \ell$) and shallow ($h_0 \ll w$) initially rectangular channel and the elastic properties of the deformable wall used in the experiments for measuring the pressure drop $\Delta p$ as a function of the flow rate $q$.}
    \begin{ruledtabular}
        \begin{tabular}{ c@{\quad} c@{\quad} c@{\quad} c@{\quad} c@{\quad}  c@{\quad}}
            %  \hline
            $h_0$ (mm)  &  $w$ (mm) & $\ell$ (mm)  & $t$ (mm) &$E$ (MPa) & $\nu$ (--)\\
            \hline
            $0.455 \pm 0.005$  & $2.9 \pm 0.06$ & $26.0 \pm 0.2$  & $0.5 \pm 0.005$ & $1.0 \pm 0.05$ & $0.48 \pm 0.1$\\
            %  \hline
        \end{tabular}
    \end{ruledtabular}
\end{table*}

\subsection{Preparation and characterization of the working fluids}

\begin{figure}[ht]
    \includegraphics[width=0.9\columnwidth]{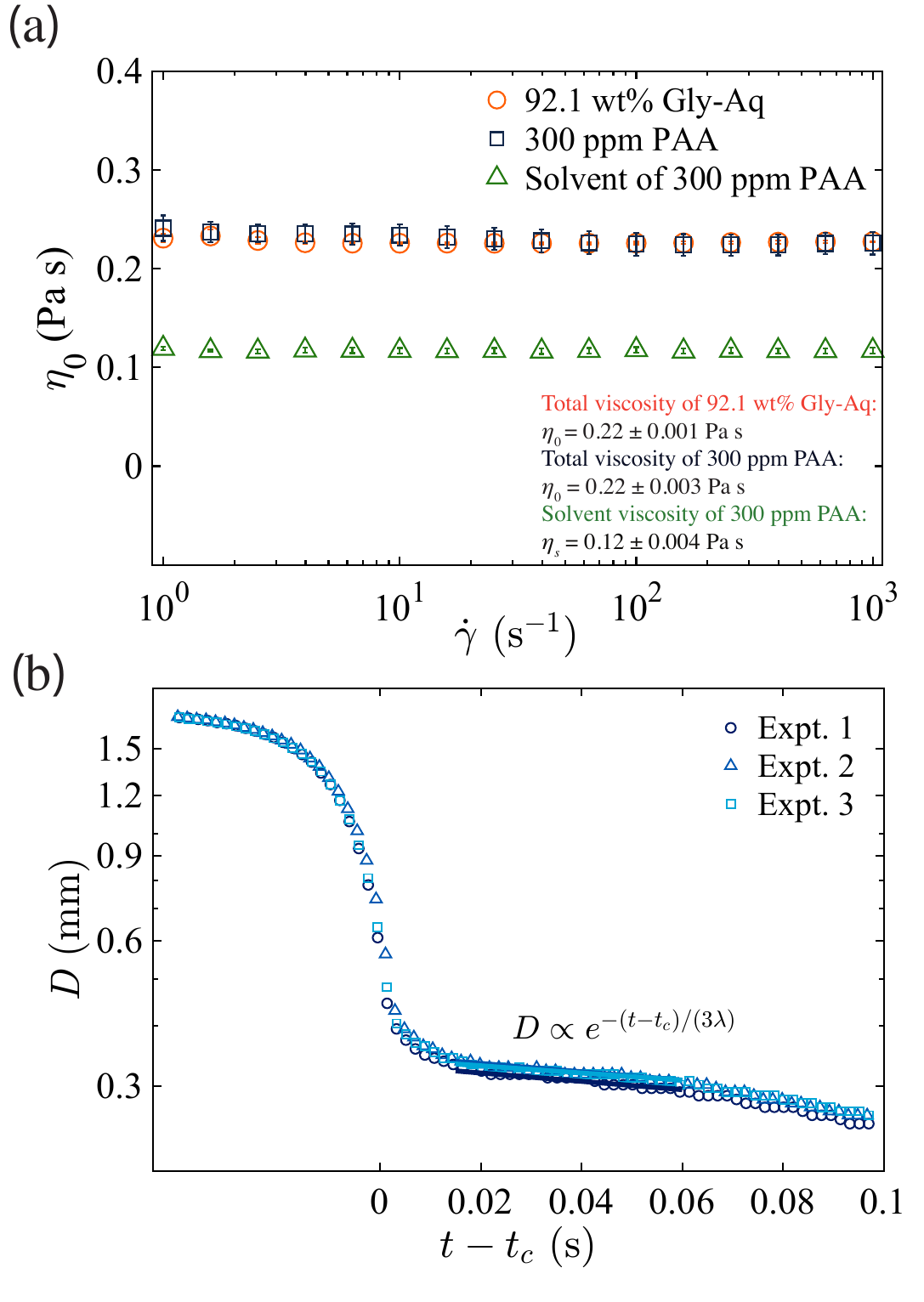}
    \caption{\label{fig2}(a) Experimental data for the shear viscosities cross several orders of magnitude of the shear rate $\dot{\gamma}$ for the Gly-Aq ($\circ$), 300 ppm PAA ($\square$) solutions and solvent of 300 ppm PAA ($\triangle$), showing negligible shear thinning. Error bars indicate the standard deviation for at least three repeated experiments under identical conditions. (b) Transient evolution of the minimum neck diameter $D$ for the 300 ppm PAA sample solutions obtained from the custom-built dripping rheometry experiments. Three independent measurements are shown. For clarity, in all cases, we only plot one point per twelve images. The extensional relaxation time $\lambda$ is extracted in the exponential thinning elasto-capillary regime by fitting. }
\end{figure}

An aqueous polymer solution with constant shear viscosity $\eta_0$ (i.e., negligible shear-thinning effects), yet exhibiting viscoelasticity (i.e., a Boger fluid) was prepared by dissolving 300 ppm of 6 MDa polyacrylamide (PAA, Polysciences), in 89 wt\% glycerol (Gly, Fisher Scientific), 10 wt\% deionized (DI) water, and 1 wt\% NaCl (Sigma-Aldrich) \cite{browne2020bistability,chen2025influence}. The Boger fluid samples were prepared by gradually dissolving the appropriate mass of powder into DI water in a cylindrical beaker, followed by continuous stirring for 24 hours to ensure complete dissolution and homogeneity. Then, we diluted the solution with 89 wt\% glycerol and added 1 wt\% NaCl, continuously stirring the mixture for another 24 hours to achieve a uniform composition. We estimate the polymer overlap concentration through the relation $C^*\approx(M_w/V)/N_A$, where $M_w$ is the molecular weight of the polymer, $V=4\pi R_g^3/3$ is the volume occupied by a single polymer molecule, and $N_A$ and $R_g$ are Avogadro's number and the radius of gyration, respectively \cite{browne2020bistability, chen2025influence}. $C^*$ defines the upper limit below which a polymer solution is considered dilute. Above this threshold, polymer chains start to enter a semi-dilute regime to overlap, interact, and may thereafter form entangled networks. The calculation yields an overlapping concentration as $C^*\approx700$ ppm \cite{ye2014interaction,lizcano2020less} and therefore, our experiments utilized a dilute polymer solution at $C/C^*\approx0.4$. For the reference Newtonian fluid as a baseline for comparison, we used an aqueous solution containing 92.1 wt\% glycerol in DI water (Gly-Aq). We performed rheological measurements of the total viscosity $\eta_0$ and solvent viscosity $\eta_s$ of the 300 ppm PAA solution, using a controlled-stress rheometer (DHR-3, TA Instruments) equipped with a cone--plate geometry (diameter = 40 mm, cone angle = $1^{\circ}$) at a controlled temperature of $25\si{\celsius}$. We observed that the 300 ppm PAA solution shows constant viscosity $\eta_0$ in a range of shear rates over three orders of magnitude ($1~\si{\per\second}\le\dot{\gamma}\le10^{3}~\si{\per\second}$), similar to that of the Gly-Aq solution in Fig.~\ref{fig2}(a).

Next, we conducted capillary breakup extensional experiments using dripping experiments \cite{bazazi2023dynamics,rajesh2022transition,turkoz2018impulsively} with a custom-built setup at room temperature $T = (25 \pm 0.5)\si{\celsius}$. In the dripping experiments, the polymer solution is gradually pumped through a vertical needle with a low flow rate ($\approx5~\si{\micro\liter/\minute}$). A fluid filament forms and subsequently thins due to the action of the capillary force. The time evolution of the minimum neck diameter $D$ of the liquid bridge is recorded. The Bond number $Bo=\rho gD^2/\sigma$ (representing the relative importance of gravitational and capillary effects), where $\sigma$ is the surface tension, is such that $Bo\ll1$ for the entire process, meaning that the gravitational force is negligible in the experiments. Fig.~\ref{fig2}(b) shows the transient evolution of $D$ for the 300 ppm PAA solution. After an initial inertial-capillary and/or viscous-capillary regime, an elasto-capillary regime emerges, wherein the polymer chains stretch and generate elastic stresses before reaching the final visco-elasto-capillary stage \cite{gaillard2024beware,dinic2019macromolecular}. In the elastic-capillary regime, a cylindrical filament thins exponentially \cite{dinic2015extensional, chandra2024elasticity} and it appears as a linear region in a semi-logarithmic plot when the shifted time is plotted along the abscissa, as shown in Fig.~\ref{fig2}(b), where $t_c$ represents the time corresponding to the transition from the Newtonian to the elasto-capillary regime. This behavior is consistent with the Oldroyd-B model, which predicts \cite{gaillard2024beware,entov1997effect,turkoz2018impulsively} $D\propto e^{-(t-t_c)/(3\lambda)}$, from which the extensional relaxation time $\lambda$ of the viscoelastic liquid was determined by fitting. We restricted the data fitting range to the early-time elasto-capillary stage, avoiding the subsequent transition to the final visco-elasto-capillary stage. We obtained a consistent value of the relaxation time $\lambda=97.6\pm0.5$ from repeated experiments. The rheological properties and composition of all the test fluids are summarized in Table~\ref{T2}.

\begin{table}[ht]
    \caption{\label{T2}Rheological properties and composition of all the test liquids.}
    \begin{ruledtabular}
        \begin{tabular}{cccccc }
            %  \hline
            Name  &  \makecell{Composition\\(\% w/w)}  & \makecell{$\eta_0$ \\(Pa s)} &\makecell{$\eta_s$ \\(Pa s)}&\makecell{$\lambda$ \\(ms)}  \\
            \hline
            300 ppm PAA  & {\makecell{PAA-0.03\\Gly-89\\DI water-10\\NaCl-1}} & \makecell{0.22\\$\pm$ 0.003} & \makecell{0.12\\$\pm$ 0.004} & \makecell{97.6 \\$\pm$ 0.5} \\
            \\
            Gly-Aq   & {\makecell{Gly-92.1\\DI water-7.9}} & \makecell{0.22\\$\pm$ 0.001} &-- & --  \\
            %  \hline
        \end{tabular}
    \end{ruledtabular}
\end{table}

\subsection{Pressure drop measurement}

To generate a steady flow in experiments, we employed a syringe pump (11 Pico Plus Elite, Harvard Apparatus) connected to the channel via Teflon tubing. The syringe pump delivered the test fluid at a constant volumetric flow rate $q$. The flow rate range was set as $0.5~\si{\milli\liter/\minute} \leq q \leq 3.5~\si{\milli\liter/\minute}$ to maintain the viscous flow regime, with low Reynolds number $Re=\rho q /(\eta_0 w) \simeq 10^{-4}$, where $\rho= 1240$ and $1190~\si{\kilogram/\meter\cubed}$ for the Gly-Aq and 300 ppm PAA solutions, respectively. The differential pressure drop $\Delta p$ along the streamwise length $\ell$ was measured using a digital pressure transducer (PX409-015GUSBH, OMEGA) connected to a digital transducer application (OMEGA) that continuously acquired data at a sampling frequency of 1000 Hz. The pressure port in the channel was positioned such that the measured length was matched with the length of the deformable portion $\ell$, as shown in Fig.~\ref{fig1}.

\section{FLOW RATE--PRESSURE DROP RELATION FOR A BOGER FLUID IN A DEFORMABLE MICROCHANNEL}

The rheological characterization thus leads us to believe that the Oldroyd-B model \cite{oldroyd1950formulation,hinch2021oldroyd} suitably captures the behavior of the 300 ppm PAA solution as a constant-viscosity polymeric liquid solution with a well-defined elastic relaxation time. Turning to the theory of viscoelastic flow (under the Oldroyd-B model) through a compliant microchannel \cite{boyko2023non}, three key dimensionless numbers govern the flow-induced deformation of a slender and shallow deformable channel:
\begin{equation}
\label{Eq.def}
    \tilde{\beta}=\frac{\eta_p}{\eta_0},\qquad
    De=\frac{\lambda}{t_r}, \qquad 
    \alpha=\frac{u_c}{h_0}.
    %\frac{(1-\nu^2)\eta_0 q \ell w^3}{2 E b^3 h_0^4}
\end{equation}
In Eq.~\eqref{Eq.def}, $\tilde{\beta}$ is the ratio of the polymeric viscosity $\eta_p=\eta_0-\eta_s$ to the total viscosity $\eta_0$ of the Boger fluid. Viscoelastic effects in the flow are quantified by $De$, which is the ratio between the relaxation time  $\lambda$ of the fluid and the residence time $t_r = \ell/v_c$ in the channel \cite{tichy1996non,anna2013non,boyko2023non} based on the characteristic axial velocity $v_c \sim q/(wh_0)$. The theory assumes that viscoelastic effects are weak, $De\ll1$. The compliance number $\alpha$ quantifies the coupling between the fluid flow and the structural deformation \cite{christov2018flow,wang2021reduced}, defined as the ratio of the characteristic deformation scale $u_c = p_c/k$ (set by the balance of the fluid's flow-rate-based pressure scale $p_c \sim \eta_0 q \ell/(h_0^3 w)$ and the bending stiffness of the wall $k \sim 2 E b^3/[(1-\nu^2)w^4]$) to the initial channel height $h_0$. 

\citet{boyko2023non} derived a set of coupled, nonlinear ordinary differential equations for the Newtonian pressure distribution in the deformable channel, $P_0(Z)$, and its $O(De)$ correction, $P_1(Z)$. Under the nondimensionalization,
\begin{equation}
    P(Z) = \frac{p(z)}{p_c},\qquad 
    Z = \frac{z}{\ell},\qquad 
    X = \frac{x}{w},
\end{equation}
and setting the outlet pressure to gauge, the governing equations are
\begin{alignat}{2}
    \frac{\mathrm{d}P_0}{\mathrm{d}Z} &= -\frac{12}{H_{e}[P_0]^{3}},\qquad & P_0(1) &= 0,
    \label{dP/dZ LO expansion}\\
    \frac{\mathrm{d}P_{1}}{\mathrm{d}Z} & = \tilde{\beta}\frac{\mathcal{S}[P_0]}{H_{e}[P_0]^{3}}, \qquad & P_1(1) &= 0,
    \label{dP/dZ FO expansion}
\end{alignat}
where the ``effective'' deformed channel height  \cite{wang2021reduced} is
\begin{equation}
    H_{e}[P_0]=\left[1+3\alpha\mathcal{I}_{1}P_0  +3\alpha^{2}\mathcal{I}_{2}P_0^{2}+\alpha^{3}\mathcal{I}_{3}P_0^{3}\right]^{1/3},
    \label{He,0}
\end{equation}
and, for convenience, we have defined
\begin{multline}
    \mathcal{S}[P_0]=\frac{\mathrm{d}P_0}{\mathrm{d}Z}\frac{36\alpha}{H_{e}[P_0]^{6}}\left[\mathcal{I}_{1}+4\alpha\mathcal{I}_{2}P_0+6\alpha^{2}\mathcal{I}_{3}P_0^{2} \right.\\
    \left. +4\alpha^{3}\mathcal{I}_{4}P_0^{3}+\alpha^{4}\mathcal{I}_{5}P_0^{4}\right]\\
    -\frac{\mathrm{d}H_{e}}{\mathrm{d}Z}\frac{108}{H_{e}[P_0]^{7}}\left[ 1+5\alpha\mathcal{I}_{1}P_0+10\alpha^{2}\mathcal{I}_{2}P_0^{2}\right.\\
    \left. +10\alpha^{3}\mathcal{I}_{3}P_0^{3}+5\alpha^{4}\mathcal{I}_{4}P_0^{4}+\alpha^{5}\mathcal{I}_{5}P_0^{5}\right] \label{S(Z)0}.
\end{multline} 
Note that $\mathcal{S}[P_0]$ and $H_{e}[P_0]$ are solely functions of $P_0=P_0(Z)$, but they depend on the parameters $\alpha$ and $\mathcal{I}_{i}$. Here, $\mathcal{I}_{i}=\int_{-1/2}^{+1/2}F(X)^{i} \,\mathrm{d}X$, $i=1,2,3,\hdots\,$ are calculated from the cross-sectional shape function of the wall displacement (induced by bending under the hydrodynamic pressure load) under the Reissner--Mindlin thick-plate theory \cite{shidhore2018static}:
\begin{equation}
    F(X)=\left(\dfrac{1}{4}-X^2\right)\left[\dfrac{2(b/w)^2}{(1-\nu)}+\left(\dfrac{1}{4}-X^2\right)\right].
    \label{eq:F_X_plate}
\end{equation}
The expressions for $\mathcal{I}_{i}=\mathcal{I}_{i}(b/w,\nu)$ are lengthy but straightforward to calculate; thus, we do not include them here. Although Eq.~\eqref{dP/dZ LO expansion} with $H_e$ from Eq.~\eqref{He,0} can be integrated analytically, it yields an implicit solution \cite{boyko2023non}. Therefore, we solve it numerically using \textsc{Matlab}'s \texttt{ode45} (with absolute and relative tolerances of $10^{-12}$). Then, we use \texttt{cumtrapz} with $\Delta Z = 2\times 10^{-3}$ to numerically evaluate $P_1(Z)$ from Eq.~\eqref{dP/dZ FO expansion}--\eqref{S(Z)0} using the numerical solution for $P_0$. Finally, $\Delta P = P_0(0) + De P_1(0)$ is the dimensionless total pressure drop, taking into account flow-induced deformation and viscoelastic effects.

\section{QUANTITATIVE COMPARISON BETWEEN EXPERIMENTS AND THEORY} % AND DISCUSSION

\begin{figure}[h]
    \includegraphics[width=0.9\columnwidth]{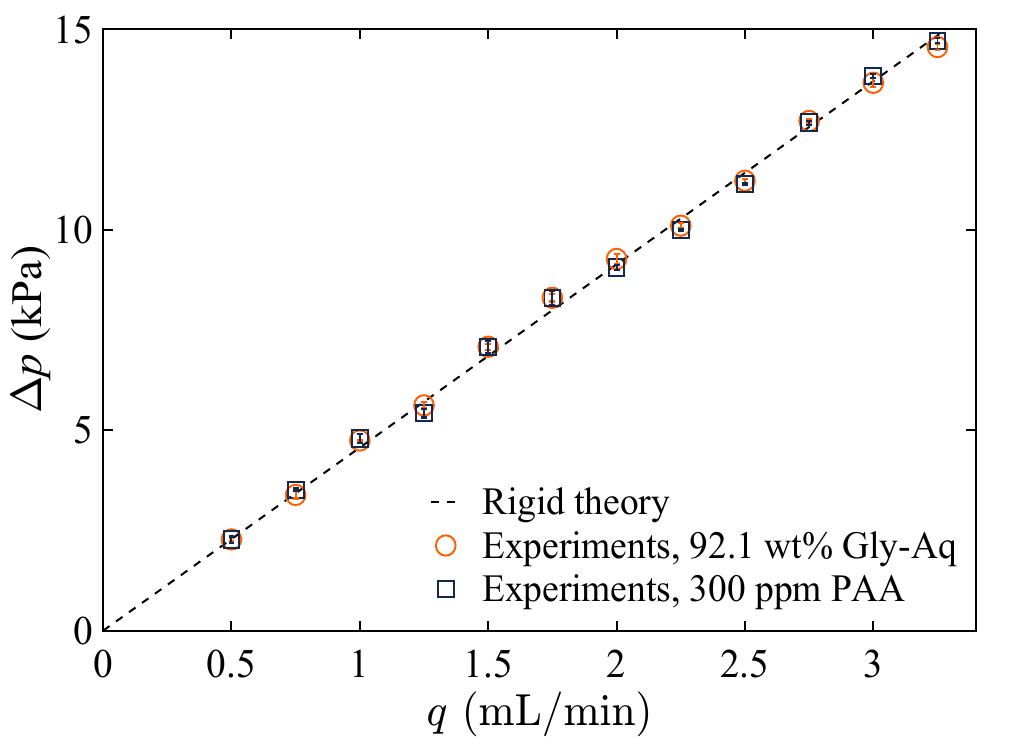}
    \caption{\label{fig_rigid}Comparison between theory, i.e., $\Delta p=12\eta_0q\ell/(wh_0^3)$ (dashed line), which holds for both Newtonian and Boger fluids in straight channels, and the experimental measurements for the flow rate--pressure drop relation in the rigid channel. Symbols represent experimental data for 300 ppm PAA ($\square$) and Gly-Aq ($\circ$) solutions. Error bars indicate the standard deviation for at least five repeated experiments under identical conditions.}
\end{figure}

Before comparing the theory to the experiments in a deformable channel, we validated the experimental setup using a rigid channel of identical dimensions. The rigid channel served as a control to ensure the accuracy and reliability of the pressure measurement system, flow rate control, and fluid injection protocols, as well as to demonstrate that there are no viscoelastic effects on the pressure drop of Boger fluids in a uniform conduit. We performed experiments with both 300 ppm PAA and Gly-Aq solutions, covering the range of flow rates used in subsequent compliant channel experiments. Fig.~\ref{fig_rigid} shows that the measured pressure drops across the rigid channel fall right on the theoretical line given by $\Delta p=12\eta_0q\ell/(wh_0^3)$, confirming that the experimental setup operated correctly under well-defined conditions and reproduces the textbook result. 

\begin{figure}
    \includegraphics[width=0.9\columnwidth]{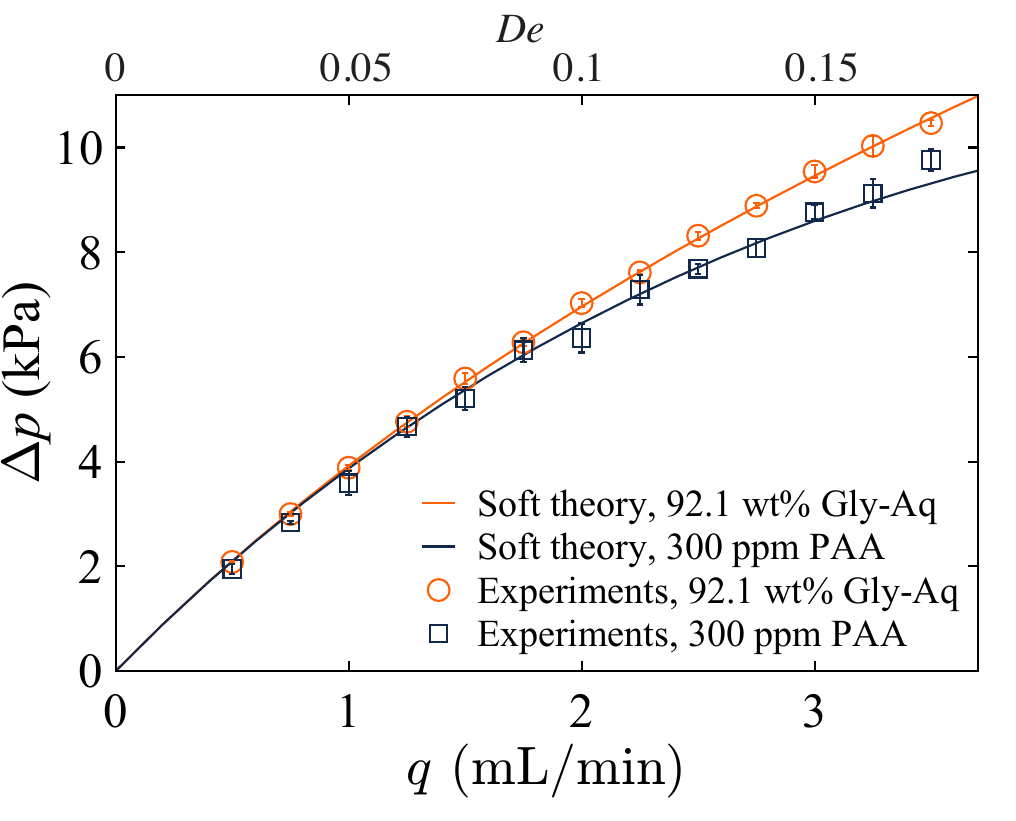}
    \caption{\label{fig_4}Comparison between theory, i.e., $\Delta p = p_0(0)$ (light curve) and $\Delta p = p_0(0) + De p_1(0)$ (dark curve) computed from Eqs.~\eqref{dP/dZ LO expansion}--\eqref{eq:F_X_plate}, and the experimental measurements (300 ppm PAA ($\square$) and Gly-Aq ($\circ$) solutions) for the flow rate--pressure drop relation in the compliant channel. Error bars indicate the standard deviation for at least five repeated experiments under identical conditions.}
\end{figure}

Fig.~\ref{fig_4} compares the theoretical predictions and experimental measurements of the $q-\Delta p$ relationship for the 300 ppm PAA and the Gly-Aq solution in the rectangular channel with a deformable top wall, over the parameter ranges $0.09\le \alpha \le0.56$ and $0.02 \le De \le0.17$. The solid curves in Fig.~\ref{fig_4} are the predicted $q-\Delta p$ relations for the deformable channel. The theoretical prediction agrees well with the experimental data for both the Boger and Newtonian fluids. Thus, the experiments have successfully verified the theory of the $q-\Delta p$ relation in a weakly viscoelastic flow of a Boger fluid in a compliant channel, making Eqs.~\eqref{dP/dZ LO expansion}--\eqref{eq:F_X_plate} ready for use in applications. 

For the Boger fluid, the theory starts to underestimate the pressure drop at flow rates $\gtrsim 3.25~\si{\milli\liter/\minute}$ ($De \gtrsim 0.16$). Of course, the theory is only valid for $De\ll1$. One reason that the pressure drop of the Boger fluid in the compliant channel deviates from the theoretical prediction may be that the lubrication-based theory posits that viscous stresses are dominant. However, the viscoelastic effects in the flow become more significant at larger $De$, and elastic normal stresses may also be generated, exerting additional forces on the deformable wall and enhancing the wall deflection, resulting in a higher pressure drop than predicted by theory \cite{boyko2024flow,boyko2022pressure}. Furthermore, we note that for $q\gtrsim 3.25~\si{\milli\liter/\minute}$, the residence time is estimated as $t_r = \ell/(q/wh_0) \lesssim 700~\si{\milli\second}$, which becomes comparable to the relaxation time of the polymeric liquid (on the order of 100 ms). In this case, there may not be sufficient time for the elastic stresses accumulated during the flow to fully relax, and the viscoelastic fluid exhibits predominantly elastic behaviors rather than viscous behaviors, as the polymers remain stretched throughout the length of the channel \cite{boyko2024flow, boyko2022pressure}. Consequently, the fluid retains the memory of its deformation history, leading to additional elastic normal stresses that resist the flow. These stresses lead to a larger pressure drop compared to what would be expected if the polymers had time to relax. 

\section{CONCLUSION AND DISCUSSION}

In this work, we investigated the interplay between purely viscoelastic rheology and wall compliance to provide a comprehensive experimental--theoretical understanding of the low-Reynolds-number flow of a Boger fluid in a rectangular channel with a deformable top wall---a prototypical geometry in microfluidic applications. We showed good agreement between theory and experiments on the steady flow rate--pressure drop relation. This quantitative comparison holds fundamental significance in soft hydraulics and non-Newtonian fluid mechanics because it clearly quantifies the viscoelastic contribution, which is only present in the deformed channels, despite the intrinsic strong coupling between wall compliance and fluid viscoelasticity. A decade ago, \citet{delg2016is} argued that microrheometry results are affected by channel deformation due to the inherent compliance of PDMS-based devices. In this study, we have precisely quantified how much the pressure drop is affected and rationalized the measurements with theory. Importantly, in this work, we performed independent rheological characterization of all fluids, leading to \emph{no fitting parameters} in the theory, with each fluid, wall, and geometric quantity being experimentally characterized. 

Thus, the experimentally benchmarked flow rate--pressure drop relation in this work provides a quantitative design principle for microfluidics and soft hydraulic systems operating with complex fluids. For instance, in an organ-on-a-chip device, it is of interest to achieve physiologically realistic shear stresses and flow rates that act on cells within the device, while taking into account the realistic compliance of the channel walls \cite{chen2024artificial}. To do so, without the need for extensive empirical calibrations or trial-and-error, the hydrodynamic resistance of the deformable channels used can be accurately estimated from the $\Delta p/q$ expressions validated in this work, taking into account wall compliance (quantified by $\alpha$) and fluid viscoelasticity (quantified by $De$). Conversely, our results provide valuable guidance for selecting materials and geometry in the design of such systems. Similarly, our work offers a systematic and quantitative framework for designing soft robotic systems, in which fluid-filled deformable channels are used for actuation or signal transduction \cite{polygerinos2017soft,coyle2018bio}, and it is necessary to precisely know the pressure to achieve a particular motion \cite{matia2023harnessing}.

Our experimental validation of the flow rate--pressure drop curve provides a reliable assessment of the adequacy of the Oldroyd-B model for describing flows of Boger fluids. However, we note that the Oldroyd-B model neglects the finite polymer chain extensibility, as well as the shear-thinning or extensional-thickening behaviors of polymeric solutions \cite{hinch2021oldroyd}. Therefore, for flow scenarios involving larger values of $De$ or more complex (shear-thinning) viscoelastic fluids, an analysis based on a more elaborate constitutive model, such as FENE-P or FENE-CR, would be required \cite{mahapatra2025viscoelastic,boyko2025pressure}. Although our experiments were deliberately designed to be in the regime where the Oldroyd-B model remains valid, future work may explore the above flow scenarios to generalize the theory and experiments to a broader class of complex fluids. In addition, elastic turbulence could occur at larger values of $De$, which is characterized by chaotic temporal and spatial velocity and stress fluctuations driven by elastic stresses \cite{li2023universal,ekanem2020signature}. The present study, with its assumptions of steady and laminar flow, cannot account for this phenomenon either.

An intriguing avenue for future research would be to investigate how the design of the pre-inlet and post-outlet sections connected to the compliant channel affects the experimental pressure drop by influencing the elastic stresses generated by the flow, as well as via flow-induced deformation of the test section. In addition, oscillatory flows are particularly relevant in biological and industrial applications \cite{dincau2020pulsatile,huh2010reconstituting}, including harnessing viscoelastic effects in the flow \cite{asghari2020oscillatory}. Therefore, oscillatory flows of Boger fluids may be the next step in the present experimental--theoretical framework for soft hydraulics with complex fluids.

\section*{ACKNOWLEDGMENTS}

We acknowledge the use of facilities and instrumentation at the Materials Research Laboratory Central Research Facilities, University of Illinois Urbana-Champaign, for the rheological experiments. S.C.\ and J.F.\ acknowledge partial support by the National Science Foundation (NSF) under grant No.~CBET 2323045 and CBET 2426809. We thank Nan Hu and Jeffrey S.\ Guasto for helpful discussions on Boger fluid preparation, Anxu Huang on microfluidic design, and Evgeniy Boyko for many fruitful discussions on viscoelastic fluid flow in deformable conduits.

\bibliography{references}

\end{document}